\documentstyle[psfig]{l-aa}
\begin{document}
\thesaurus{06         
              (11.09.3;  
               12.04.1;  
               12.07.1;  
               13.25.2)} 

\title{The ROSAT Deep Survey}

\subtitle{IV. A distant lensing cluster of galaxies with a bright arc}

   \author{G. Hasinger\inst{1} \and 
           R. Giacconi\inst{2} \and
           J.E. Gunn\inst{3} \and
           I. Lehmann\inst{1} \and  
           M. Schmidt\inst{4} \and
           D.P. Schneider \inst{5} \and
           J. Tr\"umper \inst{6} \and
           J. Wambsganss \inst{1} \and
           D. Woods\inst{1} \and
           G. Zamorani \inst{7}} 
          
   \offprints{G. Hasinger}
   
\institute{Astrophysikalisches Institut Potsdam,
              An der Sternwarte 16, D-14482 Potsdam
\and European Southern Observatory, Karl-Schwarzschild-Str. 2,
     85740 Garching bei M\"unchen, Germany
\and Princeton University Observatory, Princeton, NJ 08540, USA
\and California Institute of Technology, Pasadena, CA 91125, USA
\and Pennsylvania State University, University Park, PA 16802, USA
\and Max--Planck--Institut f\"ur extraterrestrische Physik,
     Giessenbachstr. 1, 85740 Garching bei M\"unchen, Germany
\and Osservatorio Astronomico, via Zamboni 33, I-40126 Bologna, Italy}

   \date{Received 15 September 1998; accepted 21 October 1998}
   
   \maketitle
   \markboth{Bright arc cluster}{Hasinger et al.}
   
   \begin{abstract}

An unusual double-lobed extended X-ray source (RX J105343+5735) 
is detected in the ROSAT ultra-deep HRI image of the Lockman
Hole. The angular size of the source is $1.7\times0.7$ arcmin$^2$ and
its X-ray flux is $2\times10^{-14}$ erg cm$^{-2}$ s$^{-1}$.
R-band imaging
from the Keck telescope revealed a marginal excess of galaxies
brighter than R=24.5, but Keck LRIS spectroscopy of 24 objects 
around the X-ray centroid
did not yield a significant number of concordant redshifts. The brightest
galaxy close to the centre of the eastern emission peak appears
to be a gravitationally lensed arc at z=2.570, suggesting that the
X-ray object is associated with the lens, most likely
a cluster of galaxies.
Based on a comparison of lensing surface mass density, X-ray
luminosity, morphology and galaxy magnitudes with
clusters of known distance, we argue that 
RX J105343+5735 is a cluster at a redshift around 1.
Future X-ray, ground-based optical/NIR and high resolution HST
observations of the system will be able to clarify the nature of
the object.

\end{abstract}

\section{Introduction}

X-ray surveys efficiently select cosmic X-ray sources at
distances ranging from our immediate neighbourhood to the
edge of the observable universe. By studying their properties as a
function of redshift we can probe the formation and
the evolution of structures in the universe. Surveys
at the faintest X-ray fluxes (e.g. Hasinger et al. 1998,
hereafter paper I) 
predominantly find active galactic nuclei (Schmidt
et al. 1998, hereafter paper II) often in conjunction with 
starburst regions,
which might also be responsible for the absorption
observed in the sources of the X-ray background (Fabian et al. 1998).

Clusters of galaxies are the second most abundant class of objects
in deep X-ray surveys. They are the largest bound
structures in the universe, composed of
dark matter condensations, hot X-ray emitting gas and
galaxies. X-ray radiation is an efficient means
to select clusters of galaxies, because the X-ray flux is
proportional to the square of the electron density.
The highest observed redshifts of X-ray selected clusters 
to date are $\sim$0.8-0.9 (Henry et al., 1997; Rosati et al., 1998)
and extended X-ray emission of clusters selected by other techniques
has been reported out to z=1.27 (Stanford et al., 1997;
Hattori et al., 1997).

In this letter we report on the discovery of a faint
double-lobed X-ray source, RX J105343+5735, detected in the ROSAT
ultra-deep HRI image of the Lockman Hole (paper I). 
In section 2 and 3 we present the X-ray and optical observations, and in 
section 4 we 
discuss the properties of the object in the context of what we know about 
other galaxy clusters. We use a Hubble constant 
$H_0=50~{\rm km~s}^{-1}~{\rm Mpc}^{-1}$
and a deceleration parameter $q_0=0.5$ throughout this paper. 

\section{X-ray observations} 

The ROSAT Deep Survey (RDS) project is described in paper I.
The RDS consists of a series of deep 
pointings of 207 ksec with the ROSAT PSPC and 1.31 Msec with the ROSAT HRI
in the direction of the ``Lockman Hole'', a line of sight with exceptionally
low HI column density. A catalogue of 50 X-ray sources
with fluxes brighter than $0.55\times10^{-14}~{\rm erg~cm}^{-2}~{\rm s}^{-1}$ 
has been published in paper I; the spectroscopic optical identifications 
are presented in paper II. The ultra-deep HRI pointing
covers a solid angle of $0.126~{\rm deg}^2$ inside the RDS.

In the north-east quadrant of the HRI image (see paper I) 
is RX J105343+5735, a clearly extended X-ray source with a double-lobed 
structure. In Fig.1 we show X-ray contours of this source and its 
surroundings, superposed on an optical image. The extension 
of the source is about 1.7 and 0.6 arcmin in the E-W and N-S direction, 
respectively. 
We denote the Eastern lobe (A) and the Western lobe (B); both components are 
significantly extended. Two more X-ray sources are detected at a significance
of $\sim 4\sigma$:
a possible group of galaxies at z=0.7 (C) and a QSO at z=2.572 (D).
Details for the detected X-ray sources are given in Table 1.
None of them appears in the previously published source list,
because either their off-axis angle is too large (A, B) or their 
X-ray flux too low (C, D) for the selection criteria applied in paper I.

\begin{table}[t]
\begin{center}
\caption[]{X-ray sources detected near RX J105343+5735}
\begin{tabular}{lcccc}
\hline
Source   & RA (2000) & DEC (2000) & Flux$^a$ & Type \\
\hline
A+B      & 10$^h$53$^m$43.4$^s$ & 57$^d$35$^m$21$^s$ & 2 & cluster\\
A        & 10$^h$53$^m$46.6$^s$ & 57$^d$35$^m$17$^s$ &   \\ 
B        & 10$^h$53$^m$40.1$^s$ & 57$^d$35$^m$25$^s$ &   \\ 
C        & 10$^h$53$^m$29.5$^s$ & 57$^d$35$^m$38$^s$ & 0.18 & group\\ 
D        & 10$^h$53$^m$48.2$^s$ & 57$^d$33$^m$55$^s$ & 0.17 & QSO\\ 
\hline
\end{tabular}
\end{center}
\vskip -0.3 truecm \hskip 0.3 truecm
$^a~~10^{-14}~{\rm erg~cm}^{-2}~{\rm s}^{-1}$  
\vskip -0.5 truecm
\end{table}

The extended source is detected in the PSPC pointing at an 
off-axis angle of $\sim 19~{\rm arcmin}$ and therefore is not well resolved. 
With about 590 net photons 
detected in the PSPC a coarse spectrum could be obtained. We fixed the N$_H$
value to $10^{20}~{\rm cm}^{-2}$, consistent with the 
independent determination of N$_H$ from a fit to the summed spectrum of
all resolved sources in the Lockman Hole (Hasinger et al. 1993). Using 
a Raymond-Smith model with solar abundances (Z=1) we obtain acceptable 
fits (see $\chi^2$ in Table 3) 
to the data assuming various redshifts for the source. 
The results are not significantly different for Z=0.3. 
The flux observed in the 0.5-2.0 keV band is  
$\sim2\times10^{-14}~{\rm erg~cm}^{-2}~{\rm s}^{-1}$ for all these fits.
The derived temperatures range between 1.8 and 3.5 keV, the luminosities 
(0.5-2 keV) between $10^{43}$ and $2.4\times10^{44}~{\rm erg~s}^{-1}$ 
for redshifts between 0.3 and 1.2 (see Table 3).

\begin{figure}[htp]
\centerline{\psfig{figure=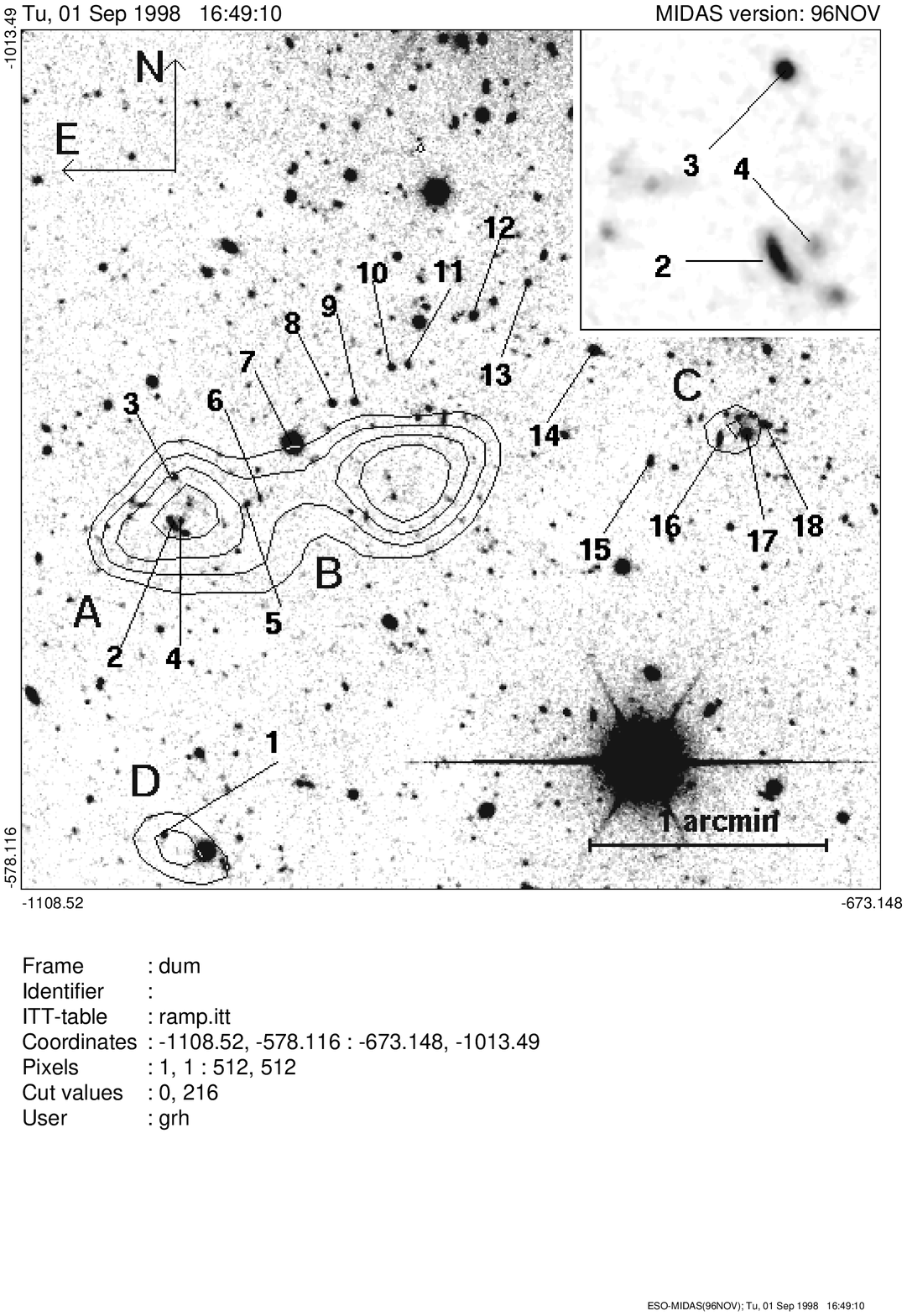,width=8.0cm,clip=}}
\caption[ ]{\small X-ray contours of a region 
about $3.6\times3.6~{\rm arcmin}^2$ in the 
ROSAT HRI ultra-deep pointing of the
Lockman Hole, superposed on a 
Keck R-band exposure. 
Numbers indicate the objects for which we have obtained
spectra. The insert in the upper right is a 
$20\times20~{\rm arcsec}^2$ zoom of the image 
close to the centre of lobe A.
An arc-like feature (\#2) is visible here.}
\vskip -0.3 truecm
\end{figure}

\begin{figure}[htp]
\centerline{\psfig{figure=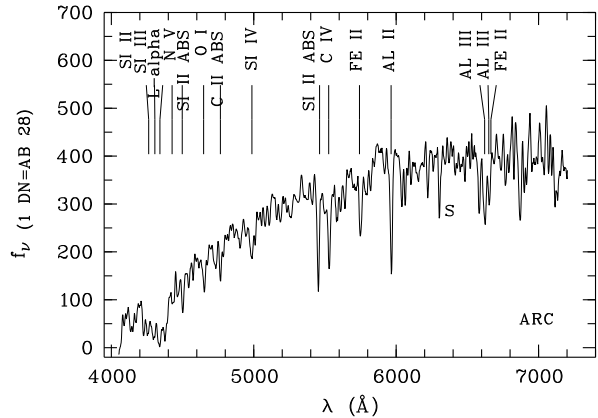,width=8.0cm,angle=-90}}
\vskip -0.5truecm
\caption[ ]{\small Keck LRIS spectrum of the arc (\#2 in Fig.1).  
The wavelengths of prominent UV metal absorption 
lines, redshifted to z=2.570 are indicated. ``S'' denotes the position of a 
strong sky line. The spectrum is typical for a high redshift starforming 
galaxy.}
\vskip -0.3 truecm
\end{figure}

\section{Optical observations}

An R-band image of the field of RX J105343+5735
was obtained on 1996 April 13 with the Low Resolution Imaging Spectrometer
(LRIS; Oke et al. 1995) on the Keck~I telescope. The exposure time was
300s and the seeing (FWHM) was 0.75". The image scale on the back-illuminated 
$2048\times2048$ Tektronix CCD is $0.215"~{\rm pixel}^{-1}$.
The CCD image was de-biased and flatfielded in the standard
fashion using IRAF routines. A flatfield was generated by
performing a median of 8 different R images, all of which
lacked bright stars and were taken over the course of the night.
We were able to flatten the resultant images to better than
$\sim1\%$.  Detection and photometric analysis of the galaxies and
stars present in the program field was done using the SExtractor
package (Bertin and Arnouts 1996). The faintest detections ($3 \sigma$)
have $R\approx25.5$.

Fig. 1 shows a section of the LRIS image underlying the HRI X-ray contours
of the field. Close to the centroid of lobe A is 
an arc-like feature (\#2). It has a length of $\sim7"$ along the major axis
and is not resolved along the minor axis. The arc has an integrated 
magnitude of R=21.4, a maximum surface brightness of 
$\mu_R=22.9$ mag arcsec$^{-2}$ 
and some asymmetry along the major axis. There is diffuse
emission around it with a surface brightness of 
$\mu_R\approx 25.5$ mag arcsec$^{-2}$ as well as some faint galaxies. 

Spectra of 24 objects in this field were acquired with a multislit mask
using LRIS on Keck~II on 1998 March 19. The conditions were 
photometric, the seeing was $\sim 1"$ and the exposure time 3600s. The slits 
were 
all 1.4" wide and some of them tilted to maximize the number of galaxies
observed. The 300 line/mm grating produced spectra in the range
3800 to 8200 \AA \ with a resolution of $\sim15$\AA.
Spectra were processed using standard MIDAS procedures for bias subtraction,
flat-field correction, optimal extraction and wavelength calibration. 
Relative flux correction and atmospheric absorption correction was done 
using the spectrophotometric standard BD +26 2606. 

Table 2 gives details about the observed galaxies using the numbering scheme
given in Fig. 1 (\#19-24 are outside the area covered by Fig. 1):  
coordinates, R magnitudes, redshifts and absolute R magnitudes 
calculated using K-corrections for the LRIS R-filter
computed assuming the spectral energy distribution for a giant elliptical 
galaxy for objects with $z<1$ and an 
assumed K-correction of 0 for the two high-redshift objects (\#1 and \#2).
Apart from four concordant redshifts around 0.7 (\#13,16,17,18), three of 
which are coincident with a discrete X-ray source (C in Fig. 1), possibly
a group of galaxies, three
galaxies at z=0.308 (\#7,11,19), scattered throughout the image, and
two galaxies at z=0.782 (\#5,8), there is no obvious 
clustering of redshifts in this sample. 

The spectrum of the brightest object 
(\#2) close to the centroid of lobe A is displayed in Fig. 2. 
It shows prominent absorption features of Ly$\alpha$ and low-ionization metal 
lines typical of the rest-frame UV emission of starburst galaxies which are 
characteristic of a young stellar population. The spectrum is very similar
to those of the UV-dropout galaxies detected by Steidel et al. (1996), which
gives us confidence in the redshift determination at $z=2.570\pm0.002$.
Its absolute magnitude
($M_R=-25.1$) indicates significant magnification and its spectrum, 
apparent magnitude (R=21.4) and redshift (z=2.570) are all very similar to 
the gravitationally 
lensed galaxy at R=21.2, z=2.515 in the cluster A2218 (Ebbels et al, 1996),
supporting the gravitational arc interpretation. Interestingly, the 
QSO (C) at z=$2.572\pm0.002$ is 
at a projected distance of only 0.6 Mpc from the arc galaxy.

\begin{table}[htp]
\begin{center}
\begin{tabular}{rccccc}
\hline
\#   & $\Delta$R.A. & $\Delta$Dec. &   R  &  z    & $M_R$\\
\hline
 1 &  32&  -91 & 22.6& 2.567 & -23.9\\
 2 &  31&  -11 & 21.4& 2.570 & -25.1\\
 3 &  31&    1 & 22.1& 0.736 & -23.4\\
 4 &  29&  -11 & 22.9&             \\
 5 &  12&   -6 & 22.9& 0.782 & -22.8\\
 6 &   9&   -4 & 23.1& 0.543:& -20.8\\
 7 &   1&   10 & 19.1& 0.308 & -22.8\\
 8 &  -9&   20 & 22.2& 0.782 & -23.5\\
 9 & -15&   21 & 22.7& 0.581 & -21.6\\
10 & -24&   30 & 22.4& 0.097 & -16.6\\
11 & -28&   30 & 22.6& 0.308 & -19.2\\
12 & -45&   43 & 21.5& 0.045 & -15.7\\
13 & -59&   52 & 22.6& 0.697 & -22.6\\
14 & -76&   35 & 21.5& 0.602:& -22.9\\
15 & -91&    7 & 22.3& 0.521 & -21.5\\
16 &-108&   13 & 22.1& 0.695 & -23.1\\
17 &-115&   14 & 21.2& 0.700 & -24.0\\
18 &-119&   17 & 22.3& 0.695 & -22.9\\
19 &-128&  109 & 20.1& 0.308 & -21.8\\
20 &-138&  126 & 22.0& 0.383 & -20.6\\
21 &-165&  136 & 20.1& 0.039:& -16.8\\
22 &-167&   92 & 20.8& 0.595 & -23.6\\
23 &-173&   19 & 21.6& 0.815:& -24.3\\
24 &-278&   44 & 19.3& 0.352 & -23.0\\
\hline
\end{tabular}
\end{center}
\caption[]{\small Parameters of galaxies with spectroscopy. \#: number of 
object 
as indicated in Fig. 1; $\Delta$R.A., $\Delta$Dec.: distance ["] relative to 
the cluster centroid centroid at R.A.$~10^h53^m43.4^s$, Dec.$~57^d35^m21^s$
(Epoch 2000); R: approximate R magnitude; z: redshift; 
$M_R$: absolute R magnitude.} 
\vskip -0.5truecm
\end{table}

\section{Discussion}

The morphology, observed flux and lensing action of RX J105343+5735 
strongly suggests that the X-ray source is associated with a moderately
luminous, distant cluster of galaxies. 
To quantify a possible excess of the projected density of galaxies
compared to the field objects, we use the estimator
discussed by Lidman and Peterson (1996, eqn. 1) which gives
a value for the cluster contrast, $\sigma_{\rm cl}$.   
Using the X-ray centroid as assumed cluster centre, we 
find a slight excess ($\sigma_{\rm cl}>2.0$) of galaxies 
with R=23.5-24.5 for a 2' diameter aperture.  The group observed
at z=0.7 (galaxies \#16, 17 and 18) shows a significant excess  
($\sigma_{\rm cl}\approx3$) of brighter galaxies ($R<23.5$). 
No obvious enhancement of galaxies is observed 
near the QSO (\#1). 

\begin{table}[t]
\begin{center}
\caption[]{Cluster parameters for assumed redshifts}
\begin{tabular}{lcccl}
\hline
Redshift &  0.3 & 0.7 & 1.2 &  \\
$kT$     &  $1.8\pm0.5$ & $2.6\pm1.1$ & $3.5\pm1.1$ & ${\rm keV}$\\
$\chi^2_{\rm red}$ &  0.49 & 0.81 & 0.89 \\
$L_X$    &  0.09& 0.66 & 2.40 & $10^{44}~{\rm erg/s}$\\
$M_L$&  0.11& 0.22 & 0.40 & $10^{14}~M_\odot$\\
$R_{\rm cD}$  & 18.8  & 22.2 & 24.9 & ${\rm mag}$\\
\hline
\end{tabular}
\end{center}
\vskip -0.5 truecm
\end{table}

Since the redshift of the lens is unknown, we have produced a series of
simple models for various values of the lens redshift
(see Table 3). For all considered redshifts the object fits onto the 
empirical luminosity-temperature relation for clusters of galaxies
(see e.g. Ebeling et al., 1996).
We estimate that the lensed arc has a radius of curvature 
$\Theta_E\approx7.5\pm2"$. 
We can estimate the lensing mass inside this radius by assuming
the arc radius corresponds to the Einstein ring radius for a mass $M_L$.
Assuming 
different redshifts we obtain the lensing masses in Table 3. Note that
this mass corresponds to the core mass of one lobe only. 
In Table 3 we also give an estimate of the expected R magnitude for the 
brightest cluster
galaxy $R_{\rm cD}$, assuming $M_R=-23$ based on the compilation of Hoessel 
\& Schneider (1985). 
We now discuss what would be the properties of the 
extended X-ray object for various assumed redshifts:

\underline{$z\approx0.3$}: Three galaxies (\#7,11,19) at this redshift are 
found, 
including the 19th magnitude galaxy (\#7) close to the symmetry axis. 
If the X-rays originate at this redshift, the X-ray luminosity would 
correspond
to a group of galaxies. The observed double structure could be similar to 
some nearby groups, which can have complicated morphologies 
(Mulchaey et al. 1996). In this case, object \#7 could be the brightest 
galaxy of the group, but no other galaxy of similar magnitude is seen around 
it. The 
detection of an arc is an argument against a group interpretation.
The lensing surface mass density would be a factor of $\sim 10$ higher than 
that typically estimated for groups (Mulchaey et al., 1996).

\underline{$z\approx0.7$}: 
Four galaxies (\#13,16,17,18) are detected at this redshift, 
but three of them are associated with source C, a very faint  
discrete X-ray source, possibly a group of galaxies ($ L_X = 5\times10^{42}$ 
erg/s). 
The X-ray luminosity would be consistent with a moderately rich cluster of 
galaxies. The elongated morphology is 
similar to other high-redshift clusters (Henry et al., 1997) and  
there is an interesting linear
arrangement between A, B and C, which could indicate that all three mass
condensations are in the same filamentary structure at z=0.7. But in this 
case A+B would have a factor of 10 higher X-ray luminosity than C, while 
C has clearly visible cluster galaxies
in the right absolute magnitude range which are not seen in A+B. Therefore
A+B would have to be a ``dark'' cluster without bright galaxies (Tucker
et al., 1995). While this is an interesting possibility, we regard it as 
unlikely.

\underline{$z>0.7$}: 
In this case all the measured properties are consistent 
with a normal, moderately rich cluster at very high redshift.
The object would be similar to the lensing cluster 2016+112 at z=1 
(Schneider et al. 1986; Hattori et al. 1997). In this scenario the 
brightest cluster galaxies might just be visible in our R-band exposure 
(see Table 3) and we would not have a chance of detecting any cluster
member within our spectroscopic limit of $R \approx 23$. 
The lensing surface mass 
density falls among those for other clusters of galaxies. In this case 
the object would be one of the highest redshift clusters with X-ray emission.
Assuming no evolution for the Rosati et al. (1998) cluster luminosity function,
we would expect 0.15 clusters in the redshift range z=1-1.5 in our survey 
volume. We regard this as the most likely possibility which can be easily 
tested with future observations. Upcoming deep surveys with AXAF and XMM should
have no problem detecting and studying a number of similar objects at high 
redshift.

In conclusion, RX J105343+5735 is an intriguing object at any of the 
redshifts we can plausibly assign to the lens and well worth follow-up
studies (e.g. NIR imaging, high-resolution imaging with HST, X-ray 
spectroscopy). In particular, it is the first detection of a gravitational 
arc which is optically brighter than any of the components of the lens.
The detection of a QSO at the same redshift as the arc indicates that an
enhanced density of background galaxies in this direction might have 
increased the likelihood to observe a lensed object. 

\begin{acknowledgements} We thank an anonymous referee for helpful
comments.
The ROSAT project is supported by the Bundesministerium
f\"ur Bildung, Forschung und Wissenschaft (BMBF), by the National
Aeronautics and Space Administration (NASA), and the Science
and Engineering Research Council (SERC). The W. M. Keck Observatory
is operated as a scientific partnership between the California
Institute of Technology, the University of California, and the
National Aeronautics and Space Administration. It was made possible
by the generous financial support of the W. M. Keck Foundation.
This work was supported by DLR grant 50 OR 9403 5 (G.H., I.L.),
National Science Foundation grant AST-95-09919 (D.P.S.) and ASI
grant ARS-96-70 (G.Z.). 
\end{acknowledgements}

\end{document}